\documentclass[fleqn]{annalen}
\usepackage{graphics}
\begin{document}
\newcommand{\volume}{8}
\newcommand{\xyear}{1999}
\newcommand{\issue}{5}
\newcommand{\recdate}{29 July 1999}
\newcommand{\revdate}{dd.mm.yyyy}
\newcommand{\revnum}{0}
\newcommand{\accdate}{dd.mm.yyyy}
\newcommand{\coeditor}{ue}
\newcommand{\firstpage}{507}
\newcommand{\lastpage}{510}
\setcounter{page}{\firstpage}
\newcommand{\keywords}{Coulomb blockade, single electron tunneling, ferromagnetic nanojunctions}
\newcommand{\PACS}{73.23.Hk, 73.40.Gk, 75.70.-i}
\newcommand{\shorttitle}{G. Micha{\l}ek et al., Theory of spin effect of single-electron tunneling}
\title{Theory of spin effect on Coulomb blockade of single-electron tunneling in ferromagnetic system}
\author{G.\ Micha{\l}ek$^{1}$, J.\ Martinek$^{1}$, J.\ Barna\'s$^{2}$, and B.\ R.\ Bu{\l}ka$^{1}$}
\newcommand{\address}
  {$^{1}$Institute of Molecular Physics, Polish Academy of Sciences, Smoluchowskiego 17,
  60-179 Pozna\'n, \hspace*{0.5mm} Poland \\
  $^{2}$Department of Physics, A. Mickiewicz University, Umultowska 85, 61-614
  Pozna\'n, \hspace*{0.5mm} Poland}
\newcommand{\email}{\tt grzechal@ifmpan.poznan.pl}
\maketitle
\begin{abstract}
Spin dependent single electron tunneling in a ferromagnetic double junction is investigated
theoretically in the limit of incoherent sequential tunneling. The junction consists of a
small nonmagnetic metallic grain with discrete energy levels, which is connected to two
ferromagnetic electrodes. We have developed a new theoretical model taking into account charge
as well as spin degrees of freedom. The model allows us to investigate new phenomena such as
spin accumulation and spin fluctuations.
\end{abstract}

\section{Introduction}

Single-electron tunneling in mesoscopic double-junctions has been extensively studied in the
last decade both experimentally and theoretically~\cite{orth}. When the central electrode
(grain) is small enough, then discrete charging of this electrode leads to Coulomb blockade of
electric current below a certain threshold voltage and to characteristic 'Coulomb staircase'
at higher voltages. The interplay of ferromagnetism and discrete charging was studied only
very recently~\cite{bar}. It was shown that spin accumulation in the grain, together with
discrete charging, is responsible for oscillations in the tunnel magnetoresistance ({\it
TMR}), i.e., in the change of junction resistance when the magnetic configuration is varied
from antiparallel to parallel alignment. In this paper we study the influence of discrete
energy levels~\cite{bee} of the grain on single-electron tunneling phenomena in ferromagnetic
junctions. The studies are performed within an approach, which is a generalization of the
orthodox method [1] for the case with many electron channels including spins.

\section{Description of the model}

We consider spin dependent transport in a double ferromagnetic nanojunction in the limit of
incoherent sequential tunneling, where the orthodox tunneling theory is applicable.
Accordingly, the resistances of the left, $R_{2\sigma}$, and right, $R_{1\sigma}$, barriers
for both spin directions ($\sigma=\uparrow$, $\downarrow$) are assumed to be much larger then
the quantum resistance $R_Q = h/2e^2$.

In the stationary state the electric current $I$ is given by
\begin{equation}
 I= e\sum_{N_{\uparrow},N_{\downarrow}}\sum_{r=\pm}
 [r\Gamma^r_{1\uparrow}(N_{\uparrow},
 N_{\downarrow})+r\Gamma^r_{1\downarrow}(N_{\uparrow},N_{\downarrow})]
 P(N_{\uparrow},N_{\downarrow}),
\label{eq1}
\end{equation}
where the probability $P(N_{\uparrow},N_{\downarrow})$ to find in the grain $N_{\uparrow}$ and
$N_{\downarrow}$ excess electrons with spin $\sigma=\uparrow$ and $\sigma=\downarrow$,
respectively, is calculated from the relevant master equation~\cite{bbmm}. In the present
model we operate in the two-dimensional space of states $(N_\uparrow, N_\downarrow)$, in
contrast to the spinless orthodox method [1] which was confined to the one-dimensional space.
The spin dependent rates of electron tunneling to ($r=+$) and off ($r=-$) the grain through
the $j$-th junction ($j=1,2$) are given by~\cite{bbmm}
\begin{eqnarray}
 \Gamma^{\pm}_{j\sigma}(N_{\uparrow},N_{\downarrow})=\frac{\Delta
 E}{e^2R_{j\sigma}} \sum_{i} \left[1+\exp
 \left(\mp\frac{E_{i\sigma}-\Delta E
 N_{\sigma}-E_F}{\frac{1}{2}k_BT}\right)\right]^{-1}\nonumber\\
 \left[1+\exp
 \left(\pm\frac{E_{i\sigma}+eV_{j}(N_{\uparrow},N_{\downarrow})\mp
 E_c-E_F}{k_BT}\right)\right]^{-1} \;,
\label{eq2}
\end{eqnarray}
where $E_F$ denotes the Fermi energy and the summation runs over discrete energy levels
$E_{i\sigma}$ of the grain, which are assumed to be equally separated with the level spacing
$\Delta E$. The formula (2) is obtained for $k_BT\approx \Delta E$, when the distribution
function of $N_{\sigma}$ among the levels of the island differs appreciably from the
Fermi-Dirac distribution function~\cite{bee}. The voltage drop
$V_j(N_{\uparrow},N_{\downarrow})$ on the $j$-th junction is given by
\begin{equation}
 V_j(N_{\uparrow},N_{\downarrow})=(-1)^jV\frac{C_1C_2}{C_jC}+\frac{e N}{C}\;,
\label{eq3}
\end{equation}
where $N=N_{\uparrow}+N_{\downarrow}$ is the total number of excess electrons on the grain,
$C_{j}$ is the capacitance of the $j$-th junction, $C$ denotes the total capacitance of the
grain $C=C_1+C_2$, while $V$, $e$ and $T$ stand for the bias voltage, electron charge and
temperature, respectively. Finally, $E_c$ is the charging energy, $E_c=e^2/2C$. We have also
assumed  that the energy relaxation time is much shorter than the time between successive
tunneling events and shorter than the spin relaxation time. Magnetic configuration of the
junction can be varied by an external magnetic field and the rotation from antiparallel to
parallel alignment is accompanied by a change in the junction resistance (in our case
$R_{1\uparrow}\to R_{1\downarrow}$, $R_{1\downarrow}\to R_{1\uparrow}$).

\section{Results and discussion}

We have calculated numerically the bias voltage dependence of the current
$I_{\uparrow\uparrow}$ and $I_{\uparrow\downarrow}$ flowing through the junction with parallel
and antiparallel alignment of the electrode magnetic moments, {\it
TMR}$=(I_{\uparrow\uparrow}-I_{\uparrow\downarrow})/I_{\uparrow\downarrow}$, spin accumulation
$<M>\equiv <N_\uparrow - N_\downarrow>$ and spin fluctuations rms$(M)=\sqrt{<M^2>-<M>^2}$. The
results are presented in Fig.1 and Fig.2 for various temperatures. For $T=2.3$ K each curve in
Fig.1 has additional steps which result from opening additional electron channels
corresponding to discrete energy levels. These steps are clearly seen only for $T \ll \Delta
E/k_B$, while for higher $T$ they are smeared out. Spin asymmetry in the tunneling rates leads
to imbalance of incoming and outgoing currents, which causes a change in the local chemical
potential and in the number of electrons of a particular spin orientation. This, in turn,
gives rise to spin accumulation (Fig.1c). One can note that in the first stage the average
$<M>$ increases with increasing bias voltage, which is due to increasing number of active
magnetic channels with increasing $V$ (see also spin fluctuations presented in Fig.1d). This
process continues until  $V \approx 20$ mV, when it is more favorable to open a new charge
channel. This, in turn, results in a reduction of the spin accumulation and spin fluctuations.
At low $T$ the shape of the spin accumulation curve for even number of electrons accumulated
in the grain is different from the shape corresponding to an odd number of electrons (this is
due to a difference in the ($N_{\uparrow}, \; N_{\downarrow}$) - space of states for the cases
of even and odd numbers). The effect of spin accumulation has a direct influence on the
character of {\it TMR} (see Fig.1b, where the small steps correspond to discrete energy levels
while the large-scale oscillations correspond to single electron charging).
\begin{figure}
\centerline{\resizebox{12.3cm}{7.45cm}{\includegraphics{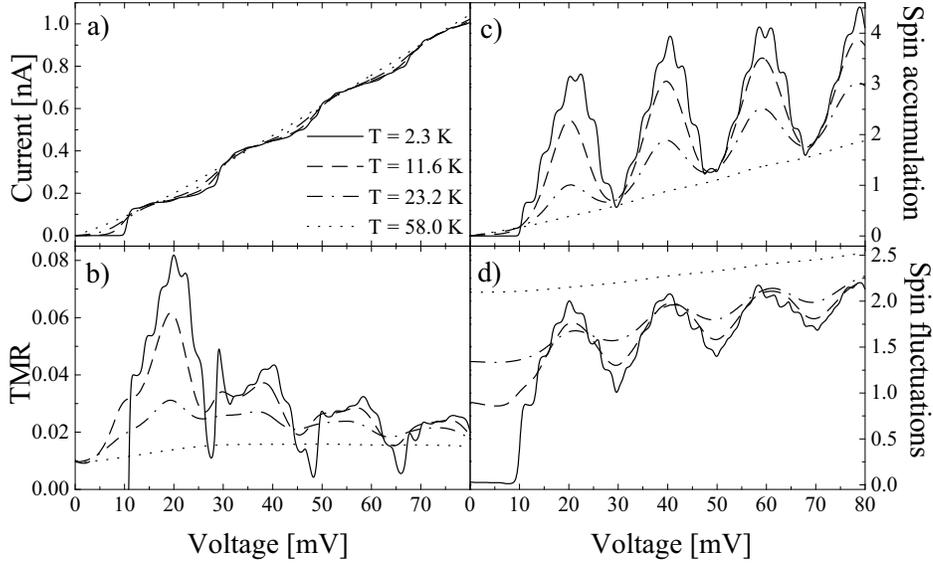}}}
 \caption{\protect Voltage dependence of the current in the antiparallel configuration (a),
 the tunnel magnetoresistance (b), the spin accumulation (c) and the spin fluctuations (d)
 for different temperatures. The parameters are: $R_{1\uparrow}=2$ M$\Omega$,
 $R_{1\downarrow}=4$ M$\Omega$, $R_{2\uparrow}=200$ M$\Omega$, $R_{2\downarrow}=100$ M$\Omega$,
 $C_1=4.3$ aF, $C_2=9$ aF, $E_c/k_B=69.9$ K and $\Delta E/k_B=34.8$ K. }
\label{fig1}
\end{figure}

The situation analysed above corresponds to a system with a large difference in the junction
resistances. In the case of a symmetrical double junction, on the other hand, the {\it I-V}
steps are not seen in the parallel configuration as shown in Fig.2a (apart from small steps
seen at low temperatures, which result from the discreteness of electronic structure).
However, the value of {\it TMR} in the ohmic limit (for high $T$) can be larger (compare Figs
1b and 2b). In the antiparallel configuration, on the other hand, the junction ceases to be
symmetrical, which leads to the oscillations in {\it TMR} with increasing $V$, seen in Fig.2b.
However, amplitude of these oscillations is quickly damped.

\section{Summary}
We presented in this paper calculations of electric current in ferromagnetic double
nanojunctions with discrete electronic structure of the central electrode. The discreteness
leads to additional structure in the {\it I-V} curves and in the voltage dependence of spin
accumulation, spin fluctuations and {\it TMR}. This fine structure is clearly seen only for
small $T$, while for higher $T$ it is smeared out. When the spin-flip relaxation time is much
longer than the time between successive tunneling events, the effect of spin accumulation
leads to {\it TMR} which oscillates with increasing $V$ (clearly seen for asymmetrical
junctions).

\vspace*{0.25cm} \baselineskip=10pt{\small \noindent The paper is supported by KBN under Grant
No. 2 P03B 075 14.

\begin{figure}
\centerline{\resizebox{12.3cm}{4.29cm}{\includegraphics{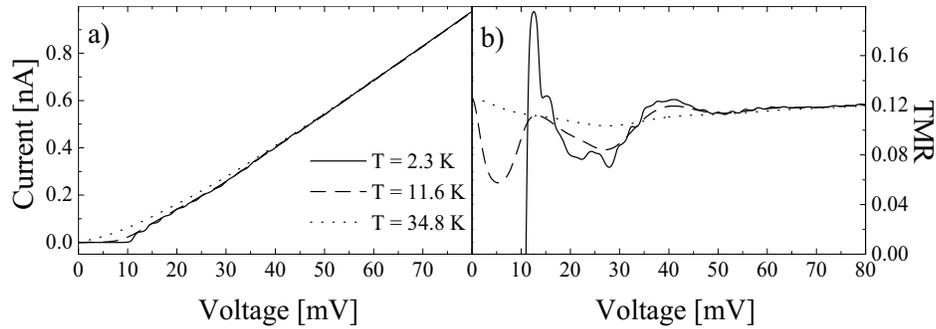}}}
 \caption{\protect Voltage characteristics of the current (a) and {\it TMR} (b) for the device with
 symmetric tunneling resistances $R_{1\uparrow}=R_{2\uparrow}=102$ M$\Omega$,
 $R_{1\downarrow}=R_{2\downarrow}=51$ M$\Omega$ in the parallel configuration. The other
 parameters are the same as in Fig.1. }
\label{fig2}
\end{figure}
\end{document}